\documentclass[amsmath,amssymb,amsfonts,twocolumn]{revtex4}
\usepackage{graphicx}

\def \beq {\begin{equation}}
\def \eeq {\end{equation}}
\def \tr {\rm Tr}

\begin{document}
\title{Quantum measurement corrections to CIDNP in photosynthetic reaction centers}
\author{Iannis K. Kominis}
\date{January 2013}
\address{Department of Physics, University of Crete, Heraklion 71103, Greece}
\begin{abstract}
Chemically induced dynamic nuclear polarization is a signature of spin order appearing in many photosynthetic reaction centers. Such polarization, significantly enhanced above thermal equilibrium, is known to result from the nuclear spin sorting inherent in the radical pair mechanism underlying long-lived charge-separated states in photosynthetic reaction centers. We will here show that the recently understood fundamental quantum dynamics of radical-ion-pair reactions open up a new and completely unexpected venue towards obtaining CIDNP signals. The fundamental decoherence mechanism inherent in the recombination process of radical pairs is shown to produce nuclear spin polarizations on the order of $10^4$ times or  more higher than the thermal equilibrium value at earth's magnetic field relevant to natural photosynthesis. This opens up the possibility of a fundamentally new exploration of the biological significance of high nuclear polarizations in photosynthesis. \end{abstract}

\maketitle
\section{Introduction}
\begin{figure}[h]
\begin{center}
\includegraphics[width=8.5 cm]{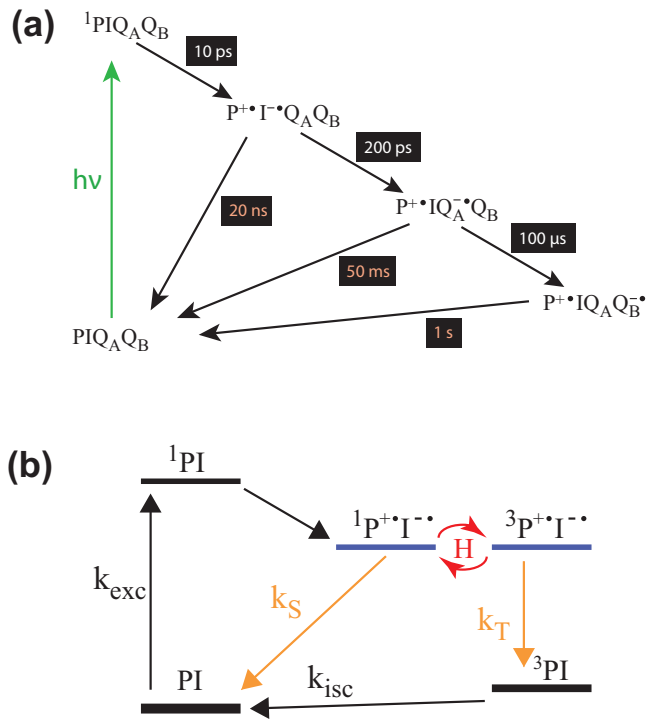}
\caption{(a) Cascade of electron transfer reactions taking place in photosynthetic reaction centers. Figure adapted from the review \cite{boxer}. (b) Primary radical-ion pair of a quinone-blocked photosynthetic reaction center. The radical-ion pair undergoes quantum dynamics due to internal magnetic interactions, embodied by the Hamiltonian ${\cal H}$, of the two unpaired electrons with each-other, the magnetic nuclei and the external magnetic field. These magnetic interactions in long-lived charge-separated states (where the long lifetime is either due to the absence of quinones or due to the large spatial separation of secondary radical-pairs) give rise to singlet-triplet conversion and spin-dependent recombination, further perplexing the dynamics of the whole reaction. Figure adapted from \cite{matysik_pnas}.}
\label{cidnp}
\end{center}
\end{figure}

The quantum dynamics of photosynthesis have, quite naturally, attracted a lot of attention recently. Understanding if and how Nature exploits quantum (de)coherence  would have tremendous scientific and technological impact. For what follows, photosynthesis can be simply visualized as a two-step process: (i) light harvesting and (ii) photochemical processing. Light harvesting molecules absorb the incident photon and guide the subsequent excitation to the photosynthetic reaction center (RC). It is there where the photochemistry takes place, transforming the initial electronic excitation in a transmembrane proton pump that further drives the life-sustaining chemistry. 
Regarding step (i), quantum coherence effects in light harvesting complexes (time scale of 1-500 fs) have been addressed theoretically and experimentally \cite{scholes,scholes_review,nalbach,mizel,ishizaki,engel,aspuru,plenio,fleming,rozzi}. We will here focus on step (ii), i.e. on reaction center quantum dynamics (time scale of  1ps - 10 $\mu$s) schematically depicted in figure \ref{cidnp}(a). 

Photosynthetic reaction centers exhibit a cascade of electron transfer reactions, starting from a photo-excited donor-acceptor dyad PI (chlorophyl-type molecules) and shelving the electron further apart  to quinones (Q$_{\rm A}$, Q$_{\rm B}$), producing a long-lived charge-separated state driving the transmembrane proton pump. In each of those steps there is a radical-ion pair formed, further perplexing the dynamics, as radical-ion pairs (RPs) exhibit intricate magnetic field effects \cite{boxer}.
Radical-ion pairs \cite{ks,steiner} are biomolecular ions created from the photo-excited dyad $^1$PI. They have two unpaired electrons (denoted by the two dots in figure \ref{cidnp}) and any number of magnetic nuclei. The magnetic nuclei of the donor and acceptor molecular ions couple to the ion's unpaired electron via the hyperfine interaction. This coupling, along with several other intra-molecule magnetic interactions, leads to singlet-triplet (S-T) mixing, i.e. a coherent oscillation of the total electronic spin, also affected by the electrons' Zeeman interaction with the external magnetic field. A spin-dependent charge recombination terminates the coherent S-T mixing and leads to the reaction products, either the original neutral dyad PI (singlet product) or a triplet intermediate ($^3$PI). As shown in figure \ref{cidnp}(b), singlet (triplet) radical pairs recombine at the singlet (triplet) recombination rate, denoted by $k_S$ ($k_T$). Triplet products may close the reaction cycle by intersystem-crossing into the singlet diamagnetic ground state at a rate $k_{isc}$.

It is apparent from figure \ref{cidnp} that when the lifetime of the charge-separated states (which increases with separation) is long enough that magnetic-field effects (with typical timescale of e.g. S-T mixing on the order of 0.1-1 $\mu$s) have time to build up, radical-pair spin dynamics get convolved into the reaction center dynamics. One aspect of this interrelation pervading a large number of photosynthetic reaction centers is 
chemically induced dynamic nuclear polarization [13-20], the enhancement (by several orders of magnitude) of the ground state (PI) nuclear spin polarization resulting from the RP spin dynamics. CIDNP has emerged as a rather universal signature of spin order in photosynthetic reaction centers \cite{matysik_PR} and its possible operational significance for photosynthesis is still an open fundamental question. CIDNP is based on nuclear spin sorting, to be shortly explained, taking place in radical-pair reactions. Regarding solid-state CIDNP in particular, three main mechanisms have so far been known to produce CIDNP signals, the three-spin-mixing (TSM) \cite{jeschke_TSM}, the differential decay (DD) and the differential relaxation (DR) mechanism \cite{JM}. The first is a coherent mechanism resulting from particular magnetic interactions within the RP, while the second is about a nuclear spin imbalance brought about by different recombination rates $k_{S}\neq k_{T}$. The last mechanism (DR) will not be relevant for the following discussion. 

\begin{figure}
\begin{center}
\includegraphics[width=8.5 cm]{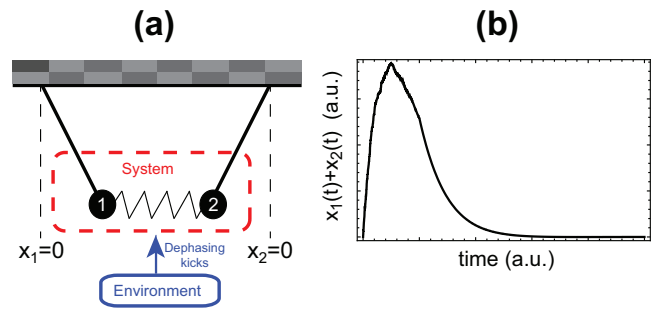}
\caption{(a) The "system" consists of $N(t)$ coupled pendulums oscillating out of phase (anti-symmetric mode). The number of such systems $N(t)$ is a decaying function of time. While they oscillate, the pendulums are perturbed by the "environment", which is here assumed to produce dephasing kicks at random times. (b) For the particular stochastic simulation shown here, every kick from the environment was taken to initialize the amplitude of pendulum 1 to unity and of pendulum 2 to zero. In the absence of the perturbation, all coupled pendulums would be continuously oscillating out of phase, and the ensemble average of the sum of their displacements away from equilibrium, $x_1+x_2$, would be constantly zero. In the presence of the dephasing kicks, however, a non-zero displacement is shown to build up, to later on decay to zero due to the total population decay.}
\label{analog}
\end{center}
\end{figure}

We will here demonstrate that there is yet another, completely unanticipated mechanism at work {\it at the fundamental quantum dynamical level of RP reactions}. It is due to the inherent quantum measurement of the RP's spin state continuously going on in the RP as part of its very charge recombination process. The decoherence caused by this "internal" quantum measurement is responsible for a new kind of nuclear spin sorting, particularly efficient at low fields, and completely unrelated to other low-field CIDNP studies \cite{jeschke_lowB}. This new mechanism is ubiquitous, as it is largely independent of the particular magnetic interactions in the radical pairs, and thus opens a new venue towards exploring the biological significance of CIDNP-induced spin order in natural (earth's field) photosynthesis. 
In the following we will (i) briefly explain the basics of CIDNP signal generation for the general reader and (ii) recapitulate the recent progress in understanding the fundamental quantum dynamics of radical-ion-pair reactions. Combining (i) and (ii), we will then explain the quantum measurement corrections to CIDNP. Before proceeding, we just mention a classical analog of the effect to be discussed, shown in figure \ref{analog}. We consider an ensemble of coupled pendulums oscillating in the 
anti-symmetric mode. The ensemble average of the sum of their displacements is $x_1+x_2=0$ at all times. However, if the "environment" perturbs the system, e.g. with random kicks, while the ensemble population decays we will observe a non-zero total displacement rising to a maximum and decaying to zero, following suit with the population decay. This displacement is the analog of the radical-pair nuclear spin to be discussed in the following. The non-zero value of the nuclear spin stems from the particular "environment" pertaining to radical-pairs, singlet-triplet dephasing, which produces a nuclear spin imbalance along the reaction. 
\section{Nuclear spin sorting}
Consider radical-ion pairs with just one nucleus coupled to the electron of e.g. the donor. At time $t=0$ the radical-ion pairs start out in the electronic singlet state, while the nuclear spins are thermally polarized, which for protons at earth's field corresponds to a polarization of about $10^{-10}$, practically taken to equal zero. Thus, in the ensemble of radical-ion pairs half will have the nuclear spin in the $|+\rangle$ state, while in the other half the nuclear spin will initially be in the $|-\rangle$ state. The local magnetic field seen by the donor electron will be the external field plus the nuclear magnetic field, while the acceptor electron will only experience the external magnetic field. It is this difference in local magnetic fields experienced by the two electrons that results in a difference in their Larmor frequencies, and hence induces the coherent singlet-triplet conversion. Consequently, in the first half radical pairs (with a $|+\rangle$ nuclear spin) the Larmor precession frequency of the  donor electron, and thus the S-T mixing frequency,  will be slightly higher than in the other half (due to the opposite contribution of the nuclear spin), hence the coherently formed triplet radical pairs will have their nuclear spins predominantly in the $|+\rangle$ state, leaving the singlet radical pairs with an opposite nuclear polarization. While the nuclear spins are thus sorted among singlet and triplet RPs, and unless there is a coherent mechanism like TSM producing net nuclear polarization, the expectation value $\langle I_{z}\rangle=\tr\{\rho I_{z}\}$ of the z-component of the single nuclear spin under consideration is still zero. Here $\rho$ is the spin density matrix of the radical pair, describing the electron and nuclear spin degrees of freedom altogether, and $z$ is the direction of the external magnetic field. 

According to the traditional understanding of the radical-pair mechanism and the concomitant CIDNP signal generation, for this nuclear spin sorting to result in a diamagnetic ground state with net nuclear polarization, and in the absence of any coherent mechanism like TSM, the differential decay mechanism (DD) should be at play in order to observe nonzero CIDNP signals. In the following we will indeed consider a magnetic Hamiltonian that {\it does not} exhibit the TSM mechanism, and we will also take $k_{S}=k_{T}$, so that the DD mechanism is not operational. Hence according to the traditional theory of RP spin dynamics, no CIDNP signal should be expected. We are then going to show that the opposite is actually true.
\subsection{Nuclear spin polarization under Hamiltonian evolution}
\begin{figure}
\begin{center}
\includegraphics[width=8.5 cm]{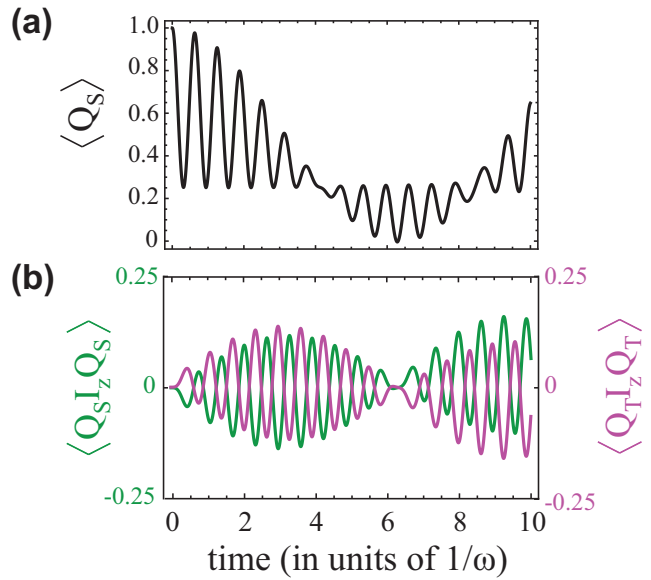}
\caption{Time evolution of (a) $\langle Q_{S}\rangle$ and (b) the nuclear spin of singlet RPs $\langle I_{z}\rangle^{S}=\langle Q_{S}I_{z}Q_{S}\rangle$ (green line) and the nuclear spin of triplet RPs $\langle I_{z}\rangle^{T}=\langle Q_{T}I_{z}Q_{T}\rangle$ (pink line), when the radical-pair spin state undergoes unceasing coherent S-T mixing driven by $d\rho/dt=-i[{\cal H},\rho]$, where ${\cal H}=A\mathbf{I}\cdot\mathbf{s}_{1}+\omega(s_{1z}+s_{2z})$. For this particular example it was $\omega=A/10$. The nuclear spin of singlet RPs is exactly out of phase with the nuclear spin of triplet RPs, so that their sum $\langle I_{z}\rangle^{S}+\langle I_{z}\rangle^{T}=0$.}
\label{traces}
\end{center}
\end{figure}

To illustrate the above considerations, we consider a radical pair with one spin-1/2 nucleus {\it isotropically} coupled to the donor electron. In figure \ref{traces}(a) we display the undisturbed spin state evolution brought about {\it just} by the magnetic Hamiltonian ${\cal H}=A\mathbf{I}\cdot\mathbf{s}_{1}+\omega(s_{1z}+s_{2z})$, where $\omega$ is the electron spin Larmor precession frequency in the external magnetic field defining the z-axis and $A$ the donor-electron hyperfine coupling with the donor nuclear spin (in units of frequency). What is plotted in figure \ref{traces}(a) is the spin state's singlet character, given by the expectation value of the singlet projection operator, $\langle Q_{S}\rangle=\tr\{\rho Q_{S}\}$, and assuming an RP with infinite lifetime (in other words so far we take the recombination rates to be $k_{S}=k_{T}=0$). What figure \ref{traces}(a) displays are the singlet-triplet oscillations brought about by ${\cal H}$. The nuclear spin, $I_z$, of singlet and triplet RPs is given by $\langle I_{z}\rangle^{S}=\langle Q_{S}I_{z}Q_{S}\rangle$ and $\langle I_{z}\rangle^{T}=\langle Q_{T}I_{z}Q_{T}\rangle$, respectively, and displayed in figure \ref{traces}(b). The previous considerations regarding nuclear spin sorting are easily visualized by noting in figure \ref{traces}(b) that although $\langle I_{z}\rangle^{S}\neq 0$ and $\langle I_{z}\rangle^{T}\neq 0$, the nuclear spin oscillations in the singlet and triplet subspace are exactly out of phase, so the net RP nuclear polarization $\langle I_{z}\rangle=\langle I_{z}\rangle^{S}+\langle I_{z}\rangle^{T}=0$. 
In case it is not obvious, we prove that $\langle I_{z}\rangle=\langle I_{z}\rangle^{S}+\langle I_{z}\rangle^{T}$ in Appendix 1. We can also show formally that for this particular Hamiltonian ${\cal H}$ it will be $\langle I_{z}\rangle=0$, as expected since ${\cal H}$ does not qualify for the TSM mechanism \cite{jeschke_TSM}. The proof can be found in Appendix 2. 
\subsection{Prediction of the traditional theory}
Now, besides the coherent RP dynamics producing the oscillations in figure \ref{traces}, which are driven by $d\rho/dt=-i[{\cal H},\rho]$ as discussed in the previous paragraph, the spin-selective charge recombination of RPs has to be accounted for. This is done by a master equation of the form $d\rho/dt=-i[{\cal H},\rho]+{\cal L}(\rho)$,
where ${\cal L}(\rho)$ is the reaction super-operator, written in terms of the singlet and triplet projection operators, $Q_S$ and $Q_T$, and the singlet and triplet recombination rates, $k_S$ and $k_T$, respectively. Recently, quantum measurement theory has been shown \cite{komPRE1,komPRE2} to fundamentally describe the reaction dynamics of RPs, resulting in a new form of ${\cal L}(\rho)$, conceptually departing from the traditional theory used until now \cite{haberkorn}.
The main physical point of departure is the fact that the very charge recombination process of RPs damps their coherent S-T mixing, i.e. it brings about an unavoidable intra-molecule decoherence in their spin state evolution. The traditional theory for the reaction operator of RPs, although successful in accounting for a large volume of experimental data, does not embody this decoherence process, neglecting which leads to several conceptual problems \cite{komCPL,komPRE3}. In any case, in the example treated above, if the nuclear spin expectation value is evaluated according to the full master equation of the traditional theory, taking $k_{S}=k_{T}$, it is found to be identically equal to zero, exactly as it is if only the Hamiltonian evolution is taken into account. This is formally shown in Appendix 3. According to the common wisdom, this should be perceived as an obvious result, since, with $k_{S}=k_{T}$ the differential decay mechanism (DD) cannot lead to any net nuclear polarization.  To recapitulate, the statement that $\langle I_{z}\rangle=0$ at all times when considering {\it just} the Hamiltonian evolution remains true {\it according to the traditional master equation of spin chemistry}, that is,  if we {\it also} take into account the recombination reaction which gradually diminishes the radical-pair population to zero.
\section{Nuclear polarization induced by singlet-triplet decoherence}
\begin{figure}
\begin{center}
\includegraphics[width=8.5 cm]{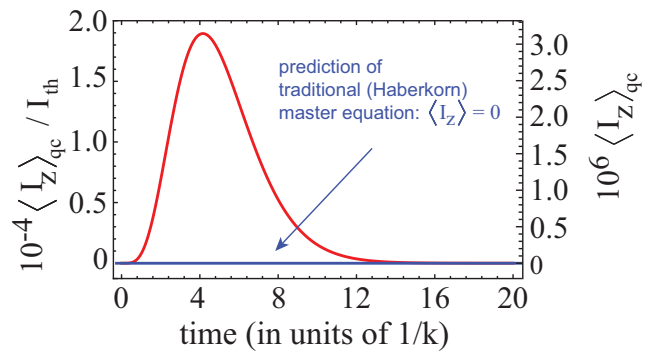}
\caption{Evolution of $\langle I_{z}\rangle$ according to the full density matrix equation of Kominis \cite{komPRE2}, with $k_{S}=k_{T}=4A$. The absolute value of the nuclear polarization is shown in the right y-axis, while its enhancement relative to the thermal equilibrium polarization, on the order of $10^4$, is shown in the left y-axis. Qualitatively the same result (only higher by about a factor of 2) is obtained from the Jones-Hore theory \cite{JH}. In contrast, the traditional theory predicts $\langle I_{z}\rangle=0$, exactly the same as the sum of the two traces in figure \ref{traces}(b) that applies to the case of no recombination reaction. Note the different time scales in this figure from figure \ref{traces}: since $k\gg\omega$, the build-up of $\langle I_{z}\rangle_{\rm qc}$ in this plot takes place at the very beginning of the S-T mixing oscillations shown in figures \ref{traces}(a) and \ref{traces}(b).}
\label{result}
\end{center}
\end{figure}

The counter-intuitive new result presented here is this:  according to the new master equation describing the fundamental quantum dynamics of RP reactions  \cite{komPRE2}, there actually is a non-zero RP nuclear spin polarization, depicted in figure \ref{result}.  A similar result, different by just a factor of 2, follows from another variant of the quantum-measurement-based stochastic Liouville equation \cite{JH}. In this particular example, this nuclear polarization is about $10^4$ times higher than the thermal equilibrium proton polarization (left y-axis of figure \ref{result}) and roughly independent of $\omega$ as long as $\omega$ is relatively small (to be further discussed later). What is the source of this new kind of CIDNP signal? We will now go beyond the particular example considered above and provide the explanation for the general case of this new channel for photochemically enhancing nuclear polarizations {\it at earth's field}: the fundamental singlet-triplet decoherence underlying the radical-pair mechanism.

To visualize this mechanism, suppose that at time $t$ the single-nuclear-spin RP ensemble is described by the density matrix $\rho$. Assuming a magnetic Hamiltonian {\it not} supporting any coherent generation of {\it net} RP nuclear polarization, as in the example considered previously, it will be $\langle I_{z}\rangle=\tr\{\rho I_{z}\}=0$. This, as already mentioned, implies that $\langle I_{z}\rangle^{S}=-\langle I_{z}\rangle^{T}$. The idea behind the new kind of CIDNP signal generation is the following. The intramolecule S-T decoherence is equivalent to some RPs being {\it projected} to either the singlet or the triplet RP manifold of states, exactly due to the continuous quantum measurement of $Q_{S}$ induced by the RP recombination dynamics, the measurement rate being $(k_{S}+k_{T})/2$ \cite{komPRE1}. The result of this measurement is either $Q_{S}=1$, i.e. the RP is projected to the singlet manifold, or $Q_{S}=0$, i.e. the RP is projected to the triplet manifold. The probability for a singlet or a triplet projection is $\langle Q_{S}\rangle$ or $\langle Q_{T}\rangle$, respectively.  Hence, at time $t+dt$ the state of the radical-pairs will be (ignoring the RPs that recombine into neutral products during $dt$) a new mixture  comprised of (a) radical pairs in the state $\rho+d\rho$, with $d\rho=-idt[{\cal H},\rho]$, plus (b) the singlet-projected radical pairs described by $Q_{S}\rho Q_{S}$, plus  (c) triplet-projected radical pairs described by $Q_{T}\rho Q_{T}$. The nuclear polarization of states (a) will still be zero, since again, the Hamiltonian ${\cal H}$ does not generate {\it net} nuclear polarization. The combined nuclear polarization of projected states (b) and (c), denoted by $\langle I_{z}\rangle^{\rm proj}$, will be the {\it weighted} sum of singlet RP nuclear polarization, $\langle I_{z}\rangle^{S}$, and triplet RP nuclear polarization, $\langle I_{z}\rangle^{T}$, the sum being weighted by the respective projection probabilities, i.e. 
\begin{equation}
\langle I_{z}\rangle^{\rm proj}=\langle Q_{S}\rangle\langle I_{z}\rangle^{S}+\langle Q_{T}\rangle\langle I_{z}\rangle^{T}
\end{equation}
Taking into account that  $\langle I_{z}\rangle^{T}=-\langle I_{z}\rangle^{S}$ and using the completeness relation $\langle Q_{T}\rangle=1-\langle Q_{S}\rangle$, it follows that 
\begin{equation}
\langle I_{z}\rangle^{\rm proj}=\langle I_{z}\rangle^{S}\Big(2\langle Q_{S}\rangle-1\Big)\label{eq:Izpr}
\end{equation}
Thus, if $\langle Q_{S}\rangle\neq\langle Q_{T}\rangle\neq 1/2$, more (when $\langle Q_{S}\rangle>1/2$) or less (when $\langle Q_{S}\rangle<1/2$) projections will take place to the singlet state than to the triplet, and $\langle I_{z}\rangle^{\rm proj}_{t}$ will be non-zero, if of course $\langle I_{z}\rangle^{S}$ is non-zero. 

To recapitulate the physics of this new effect, the singlet-triplet dephasing inherent in the radical-pair quantum dynamic evolution randomly projects radical-pairs into either the singlet or the triplet subspace. The projection probability at time $t$ depends on the singlet (or triplet) character of the electronic spin state, hence the projections will be asymmetric if $\langle Q_{S}\rangle\neq 1/2$. Nuclear spin sorting, i.e. the fact that $\langle I_{z}\rangle^{S}=-\langle I_{z}\rangle^{T}\neq 0$, combined with the asymmetric projections produces a non-zero nuclear spin expectation value.
\section{Estimate of the quantum measurement corrections to CIDNP for the general case}
We can now theoretically estimate the quantum measurement corrections to CIDNP, to be denoted by $\langle I_{z}\rangle_{\rm qc}$, shown in figure \ref{result}. Letting $k=k_{S}=k_{T}$, the rate of singlet or triplet projections (the "quantum measurement" rate) is just $(k_{S}+k_{T})/2=k$. We will focus on low magnetic fields and assume, as is the case most often, that $k>A,\omega$. Towards the estimate, consider the following three steps: (1) As the RPs start out in the singlet state, and the S-T mixing frequency $\Omega\ll k$, in the short reaction time of $1/k\ll 1/\Omega$ the singlet character will not have enough time to change appreciably, so it will be $\langle Q_{S}\rangle\approx 1$. This means that the RPs will be projected predominantly to the singlet state. (2) The electron spin of the donor molecule, which contains the one and only nuclear spin of the RP under consideration, will precess at frequency $|\omega\pm A|$ when the nuclear spin is in the $|\pm\rangle$ state. So at time $t$, the number of singlet RPs with the nuclear spin in the $|\pm\rangle$ state will be $N_{\pm}\approx\cos[(\omega\pm A)t]^2$. The nuclear polarization of singlet RPs at time $t=1/k$ will thus be $\langle I_{z}\rangle^{S}=(N_{+}-N_{-})/(N_{+}+N_{-})$, easily seen to be
\begin{equation}
\langle I_{z}\rangle^{S}\approx  -{{\omega A}\over k^{2}}
\end{equation}
And (3), S-T projections taking place at the rate $k$ will be a source term for $d\langle I_{z}\rangle_{\rm qc}/dt$. The  source term, proportional to $\langle I_{z}\rangle^{\rm proj}$, starts at $t=0$ from zero and oscillates at a rate $\Omega$ while it is drained at a rate $k$, thus $d\langle I_{z}\rangle_{\rm qc}/dt=k\langle I_{z}\rangle^{\rm proj}e^{-kt}\sin\Omega t$, yielding at time $t\approx 1/k$ that 
$\langle I_{z}\rangle_{\rm qc}\approx -\langle I_{z}\rangle^{\rm proj}\Omega^{2}/k^{2}$. Putting everything together and using equation (\ref{eq:Izpr}), we finally arrive at 
\begin{equation}
\langle I_{z}\rangle_{\rm qc}\approx {{\omega\Omega^{2} A}\over k^{4}}\label{estm}
\end{equation}
For the particular example depicted in figure \ref{traces} it is $\Omega\approx A/10$, hence according to the estimate of equation (\ref{estm}) it is $\langle I_{z}\rangle_{\rm qc}\approx 4\times 10^{-6}$, in good agreement with the exact value shown in the right y-axis of figure \ref{result}. Furthermore, the result of equation (\ref{estm}) can be recast in terms of the thermal polarization $I_{\rm th}=\hbar\omega/4\gamma k_{B}T$ (the expression for $I_{\rm th}$ is valid for $\hbar\omega\ll\gamma k_{B}T$), where $\omega=\gamma_{e}B$ is the electron Larmor frequency in the external magnetic field $B$, with $\gamma_{e}=2\pi\times$ 2.8MHz/G being the electron gyromagnetic ratio and $\gamma=|\mu_{e}/\mu_{p}|=658.5$ is the ratio of electron-to-proton magnetic moment. The enhancement factor $\langle I_{z}\rangle_{\rm qc}/I_{\rm th}$ is thus seen to be independent of $\omega$ and equal to
\begin{equation}
{{\langle I_{z}\rangle_{\rm qc}}\over {I_{\rm th}}}\approx 10^{3} {{\Big(\Omega/0.01~{\rm ns}^{-1}\Big)^2\Big(A/0.1~{\rm ns}^{-1}\Big)}\over {\Big(k/1~{\rm ns}^{-1}\Big)^{4}}}
\end{equation}
For ease of use, in the previous expression we provided the enhancement factor in terms of the S-T mixing frequency $\Omega$ (in units of $0.01~{\rm ns}^{-1}$), the hyperfine coupling $A$ (in units of  $0.1~{\rm ns}^{-1}$) and the recombination rate $k$ (in units of $1~{\rm ns}^{-1}$).
For typical hyperfine couplings of about 10 Gauss and recombination times on the order of 1 ns, it is seen that the quantum measurement corrections to chemically induced dynamic nuclear polarization amount to a significant enhancement of at least four orders of magnitude over the thermal polarization. 

We note for the reader who would worry about angular momentum conservation that the RP's nuclear spin polarization is exactly balanced by an opposite and equal electronic spin polarization, i.e. the sum $\langle s_{1z}+s_{2z}+I_{z}\rangle=0$ at all times, as it should be, since it is zero at time $t=0$ (the initial state is $\rho_{0}=Q_{S}/\tr\{Q_{S}\}$, which is a singlet state with zero nuclear polarization). The non-zero electronic spin polarization necessarily involves the admixture of the $|T_{+}\rangle=|\uparrow\uparrow\rangle$ and $|T_{-}\rangle=|\downarrow\downarrow\rangle$ triplet states, therefore this new CIDNP mechanism is a low-magnetic-field effect, because at high fields the $|T_{\pm}\rangle$ states are out of reach from the singlet state $|S\rangle=(|\uparrow\downarrow\rangle-|\downarrow\uparrow\rangle)/\sqrt{2}$. How low a magnetic field depends on the recombination rate $k$, since the triplet RP decay broadens the triplet energy levels by about $k$, hence the mechanism is appreciable for magnetic fields $B$ such that $B<k/\gamma_{e}$, i.e. when the Zeeman energy separation of $|T_{\pm}\rangle$ is within the broadening $k$. For $k\approx 1~{\rm ns}^{-1}$, the effect is appreciable for fields $B\leq 50~{\rm G}$.
\section{Conclusions}
In closing, the following comments should be made:\newline
{\bf (1)} We stress that the new CIDNP signal is a fundamental phenomenon attributed to the inherent quantum dynamics of radical-ion pairs. \newline
{\bf (2)} Kaptein's rules provide a compact relation of the main features of the CIDNP signal to the system's parameters. For this case, the sign of the effect, i.e. whether it is emissive (+) or absorptive (-) is the same as the sign of the hyperfine coupling $A$.\newline
{\bf (3)} The new CIDNP effect reported here does not require any particular combination or fine-tuning of magnetic interactions, like the coherent TSM low-field effect analyzed in \cite{jeschke_lowB}, but is rather general and shows up whenever there is a non-zero nuclear spin polarization of singlet RPs, $\langle I_{z}\rangle^{S}$, as in figure \ref{traces}(b). The rather general requirement for this to happen is that the oscillation frequencies $\omega_{m}-\omega_{n}$  of the non-zero matrix elements $[Q_{S}]_{mn}=\langle m|Q_{S}|n\rangle$ must have terms linear in the hyperfine couplings. Here $|n\rangle$ and $\omega_{n}$ are the eigenvectors and eigenvalues of ${\cal H}$, with $n=1,...,4M$, where $M$ is the nuclear spin multiplicity of the radical-pair. \newline
{\bf (4)} Regarding experimental detection of the new CIDNP effect we note the following. While NMR measurements at low fields suffer loss of sensitivity, recent methods of ultra-sensitive magnetometry using SQUIDs \cite{squidNMR1,squidNMR2} or atomic magnetometers \cite{atomicNMR1,atomicNMR2} have demonstrated the ability to sensitively detect NMR signals at near-zero magnetic fields. To provide a rough estimate of the magnetic field $B_{n}$ produced by the polarized nuclear spins, we note that in the solid phase the concentration of reaction centers can be as large as 1~mM. Assuming just a single proton spin per RC (which will heavily underestimate the magnitude of the effect), choosing a magnetic field of 1 G and taking into account that the considered enhancement will boost the thermal nuclear polarization by a factor $10^4$, we find that the magnetic field just outside a spherical volume will be $B_{n}\approx 10^4P_{\rm th}\mu_{p}\mu_{0}[{\rm RC}]$ where $[{\rm RC}]$ is the reaction center concentration. At 1 G the thermal polarization is $P_{\rm th}\approx 10^{-10}$, hence $B_{n}\approx 20~{\rm fT}$, already 200 times higher than what state-of-the-art atomic magnetometers can do \cite{subfemtomagn}. In fact, this seems to be yet another exciting application of ultra-sensitive magnetometry in the biological realm \cite{meg,plants}.\newline
{\bf (5)} Although we considered the special case $k_{S}=k_{T}$, the new kind of CIDNP signal we predict will ubiquitously affect all sorts of radical pairs, including of course the ones with asymmetric recombination rates $k_{S}\neq k_{T}$, which predominantly occur in photosynthetic reaction centers \cite{matysik_PR}. We have chosen to deal with the special case $k_{S}=k_{T}$ and a magnetic Hamiltonian unable to support coherent generation of CIDNP signals in order to be able to focus exclusively on this fundamentally new CIDNP process. Establishing how this mechanism attributed to S-T decoherence affects, qualitatively and quantitatively, the cases where {\it there exists} a CIDNP signal according to the mechanism's traditional understanding (e.g TSM, DD or DR scenarios) is more involved an exercise that will be undertaken elsewhere.\newline
{\bf (6)} We have considered the CIDNP signal in the RP state, and not in the diamagnetic ground state of the molecule in a closed reaction cycle as the one shown in figure \ref{cidnp}(b). In other words in figure \ref{result} we depicted the transient nuclear polarization of the radical-ion pairs in between their creation at time $t=0$ and the time their population has decayed to zero due to charge recombination. However, it is easily seen that the same results hold true for the diamagnetic ground state in the steady state under continuous excitation (i.e. continuous illumination) and a closed reaction cycle, as long as the intersystem crossing rate $k_{isc}$ (see figure \ref{cidnp}b) is fast enough.

Concluding, this is the first step of a very promising research program, namely what are the particular repercussions of the recently unraveled fundamental quantum dynamics of the radical pair mechanism for  understanding the dynamics of photosynthetic reaction centers. We have extended the fundamental understanding of CIDNP by pointing to a new and qualitatively different way to obtaining enhanced nuclear spin polarization. Interestingly, this polarization {\it naturally} turns out to be significant at low magnetic fields pertinent to natural photosynthesis. It stems from the random singlet or triplet projections {\it naturally} taking place during the quantum evolution of radical pairs and leading to spin decoherence. Thus, an intriguing question raised is if and how the quantum mechanical process of spin decoherence affects the biological workings of photosynthetic reaction centers. In other words, what is the biological significance of earth-field CIDNP produced by S-T decoherence? 

\section*{Appendix 1}
Since $I_{z}$ commutes with the electron spins $\mathbf{s}_{1}$ and $\mathbf{s}_{2}$, it also commutes with $Q_S=1/4-\mathbf{s}_{1}\cdot\mathbf{s}_{2}$ and with $Q_T=1-Q_S$. Thus, since the two projection operators are orthogonal, i.e. $Q_{S}Q_{T}=Q_{T}Q_{S}=0$, it will be $Q_{S}I_{z}Q_{T}=Q_{T}I_{z}Q_{S}=0$. Furthermore, since $Q_{S}+Q_{T}=1$, it will be $\tr\{\rho I_{z}\}=\tr\{\rho(Q_{S}+Q_{T})I_{z}(Q_{S}+Q_{T})\}$, and by using the previous relations it follows that indeed $\langle I_{z}\rangle=\langle I_{z}\rangle^{S}+\langle I_{z}\rangle^{T}$.
\section*{Appendix 2}
Since $I_{z}$ commutes with $\mathbf{s}_{1}$ and $[I_{z},I_{x}]=iI_{y}, [I_{z},I_{y}]=-iI_{x}$, it follows that $[I_{z},{\cal H}]=iA(s_{1x}I_{y}-s_{1y}I_{x})$. Then, Heisenberg's equation of motion $dI_{z}/dt=i[I_{z},{\cal H}]$ leads to 
\begin{equation}
d\langle I_{z}\rangle/dt=-A\langle s_{1x}I_{y}-s_{1y}I_{x}\rangle\label{eq:dIzdt}
\end{equation}
The most general pure spin state of the RP can be written as $|\psi\rangle=\sum_{i=S,T_{0},T_{++},T_{--}}|i\rangle|\chi_{i}\rangle$, where $|i\rangle$ are the four electron-spin S-T basis states and $|\chi_{i}\rangle$ is a nuclear spin state. It is straightforward to show that $\langle\psi|s_{1x}I_{y}-s_{1y}I_{x}|\psi\rangle=0$, therefore $d\langle I_{z}\rangle/dt=0$ at all times. The same holds true for any mixture of pure states. Since at time $t=0$ we have an RP mixture with zero nuclear polarization, at all times it will be $\langle I_{z}\rangle=\langle[{\cal H},I_{z}]\rangle=0$.  
\section*{Appendix 3}
The traditional master equation is $d\rho/dt=-i[{\cal H},\rho]-k_{S}(Q_{S}\rho+\rho Q_{S})/2-k_{T}(Q_{T}\rho+\rho Q_{T})/2$. For the case under study $k_{S}=k_{T}=k$, and taking into account that $Q_{S}+Q_{T}=1$, the traditional master equation becomes $d\rho/dt=-i[{\cal H},\rho]-k\rho$. After multiplying with $I_{z}$ and taking the trace, it follows that according to the traditional theory it will be $d\langle I_{z}\rangle/dt=-i\tr\{[{\cal H},\rho]I_{z}\}-k\langle I_{z}\rangle$. But $\tr\{[{\cal H},\rho]I_{z}\}=-\tr\{[{\cal H},I_{z}]\rho\}=-\langle[{\cal H},I_{z}]\rangle$, which was shown to equal zero in Appendix 2. Therefore $d\langle I_{z}\rangle/dt=-k\langle I_{z}\rangle$. Since at $t=0$ it is $\langle I_{z}\rangle=0$, the same is true at all times.

\section*{Acknowledgements}
I warmly acknowledge helpful comments from Profs. Gunnar Jeschke and Ulrich Steiner, and several comments clarifying CIDNP-related issues from Prof. J\"{o}rg Matysik.


\end{document}